\documentclass{ws-procs11x85}
\usepackage{ws-procs-thm}           
\usepackage[draft]{fixme}
\fxsetup{inline,nomargin,theme=color}

\begin{document}

\title{Intimate Partner Violence and Injury Prediction From Radiology Reports}

\author{Irene Y. Chen$^{1}$\footnote{Corresponding author email: iychen@mit.edu.}, Emily Alsentzer$^2$, Hyesun Park$^3$, Richard Thomas$^4$, Babina Gosangi$^5$, \\Rahul Gujrathi$^3$, and Bharti Khurana$^{3,6}$}

\address{$^1$Electrical Engineering and Computer Science, Massachusetts Institute of Technology,\\
Cambridge, MA 02139, USA}

\address{$^2$Health Sciences and Technology, Massachusetts Institute of Technology,\\Cambridge, MA 02139, USA}

\address{$^3$Department of Radiology, Brigham and Women’s Hospital, Boston, MA 02115, USA}

\address{$^4$Department of Radiology, Lahey Health Medical Center, Burlington, MA 01805, USA}

\address{$^5$Department of Radiology, Yale University, New Haven, CT 06510, USA}

\address{$^6$Department of Radiology, Harvard Medical School, Boston, MA 02115, USA}

\begin{abstract}
Intimate partner violence (IPV) is an urgent, prevalent, and under-detected public health issue. We present machine learning models to assess patients for IPV and injury. We train the predictive algorithms on radiology reports with 1) IPV labels based on entry to a violence prevention program and 2) injury labels provided by emergency radiology fellowship-trained physicians. Our dataset includes 34,642 radiology reports and 1479 patients of IPV victims and control patients. Our best model predicts IPV a median of 3.08 years before violence prevention program entry with a sensitivity of 64\% and a specificity of 95\%. We conduct error analysis to determine for which patients our model has especially high or low performance and discuss next steps for a deployed clinical risk model. 
\end{abstract}


\copyrightinfo{\copyright\ 2020 The Authors. Open Access chapter published by World Scientific Publishing Company and distributed under the terms of the Creative Commons Attribution Non-Commercial (CC BY-NC) 4.0 License.}
\section{Introduction}


Intimate partner violence (IPV) is defined as physical, sexual, psychological, or economic violence that occurs between former or current intimate partners. While men can also be affected, IPV is a gendered phenomenon largely perpetrated against women by male partners.~\cite{fulu2013prevalence} The Centers for Disease Control report that more than 1 in 3 women, and 1 in 10 men in the U.S. will experience physical violence, sexual violence, psychological violence, and/or stalking by an intimate partner during their lifetime.~\cite{black2011national} IPV victims have a greater risk of health problems including higher rates of mental health illnesses, chronic pain, reproductive difficulties, and generally poorer health.\cite{campbell2002health, tollestrup1999health, ellsberg2008intimate} According to the United Nations, half of the women who are intentionally killed globally are killed by their intimate partners or family members.~\cite{united2018global} It is essential to detect IPV victims early to provide timely intervention.

Healthcare providers have the opportunity to screen patients for IPV, but several barriers at both patient and provider levels limit the effectiveness. 
IPV victims often seek treatment within healthcare settings;~\cite{wisner1999intimate} however, despite its high prevalence, IPV is substantially underdiagnosed due to underreporting of violence by the victim to health care providers. Because IPV victims generally do not present with obvious trauma, even in emergency departments,~\cite{dearwater1998prevalence} they do not readily receive IPV-specific resources. 

Imaging studies provide an objective measurement of patient status, especially for vulnerable individuals who are not forthcoming.~\cite{russo2019imaging} In a prior observational study, researchers identified IPV-related injury patterns including soft-tissue and musculoskeletal injuries from imaging studies of victims who visited the emergency department. They also found that IPV victims receive more radiology studies than a comparable control cohort.~\cite{george2019radiologic} 

In this work, we present algorithms to predict IPV and injury from radiology reports. We predict IPV from a dataset of 24,131 radiology reports from 262 IPV victims who enrolled into a violence prevention support program and 794 controls from the same hospital who were age and sex-matched based on a subset of the IPV victims. We demonstrate strong quantitative results with our best model achieves a mean area under the received operator curve (AUC) of 0.852. With a sensitivity of 64\% and a specificity of 95\%, we are able to predict IPV a median of 3.08 years in advance of entry into the violence prevent support program. To better detect severe forms of IPV, we predict injury from a dataset of radiology reports from only IPV victims with labels from four emergency radiology fellowship-trained radiologists. Our best model achieves a mean AUC of 0.887. 

We analyze our models for validity and usability. 
Because IPV can manifest differently across race,~\cite{lipsky2006role} gender,~\cite{tjaden2000prevalence} age,~\cite{rennison2001intimate} and marital status,~\cite{salomon2002relationship} we present error analysis comparing accuracy, sensitivity, and specificity across these groups using demographic information extracted from the clinical record. 
As IPV continues to affect vulnerable individuals---especially in times of great crisis~\cite{van2020covid,gosangi2020exacerbation}---we demonstrate how automated predictive algorithms can be used to identify patients at high risk of IPV and injury.


\section{Related Work}

\subsection{Intimate partner violence}

Early detection in IPV is critical to facilitate early intervention in the cycle of abuse, thereby preventing worsening health conditions,~\cite{campbell2002health, tollestrup1999health, ellsberg2008intimate} life threatening injuries, and potentially homicides.~\cite{stockl2013global} The main obstacle to early intervention is underreporting by the patient due to variety of factors including shame, economic dependency, or lack of trust in healthcare providers.~\cite{hien2009interpersonal}  Automated screening can help physicians identify high risk individuals---potentially from radiology studies~\cite{khurana2020making}, substance abuse disorders~\cite{kraanen2014prediction}, or other clinical data---and intervene quickly. Prior work has focused on analyzing associative patterns among IPV victims.~\cite{george2019radiologic} To our knowledge, we present the first work to present an algorithm for IPV and injury prediction.




\subsection{Clinical prediction}

Machine learning methods can assess patients and other individuals for different levels of risk to allocate resources and improve clinical workflows.~\cite{ghassemi2020review, ghassemi2019practical} The strength of machine learning lies in its ability to learn latent patterns from observational data and make robust predictions on new and previously unseen patients. Researchers have shown promising results about the use of machine learning on chronic diseases like diabetes~\cite{razavian2015population}, diagnosis from radiology reports~\cite{kehl2019assessment}, rare conditions like preterm infant illnesses,~\cite{saria2010integration} and public health concerns like child welfare.~\cite{chouldechova2018case, brown2019toward} In particular, supervised learning models excel in structured settings with large datasets and clearly defined labels, e.g. radiology report text and whether the patient ultimate enters a violence prevention program.

\subsection{Natural language processing}

%
Natural language processing (NLP) techniques can extract information from unstructured text~\cite{wu2020deep}. In healthcare settings, researchers have leveraged NLP on clinical text such as nursing notes, discharge summaries, and radiology and pathology reports for  disease surveillance~\cite{kehl2019assessment,gao2018hierarchical}, cohort creation~\cite{chen2019clinical,afzal2018natural}, prediction of adverse events~\cite{poulin2014predicting,huang2019clinicalbert,rumshisky2016predicting}, and diagnosis~\cite{pham2014natural,boag2018s}.

%
A promising new area of natural language processing research is the use of contextual word embeddings. Whereas traditional approaches represent text as a non-sequential bag of words or a sequence of static word embeddings, more recent approaches construct unique representations for each word (or sub-word) depending on its surrounding context. For instance, the abbreviation ``MS" may refer to mitral stenosis or multiple sclerosis depending on the surrounding context. BERT~\cite{devlin2018bert}, RoBERTa~\cite{liu2019roberta}, AlBERT~\cite{lan2019albert}, and numerous other recent models are pretrained on large amounts of text using language modelling objectives and then fine-tuned on a smaller task-specific dataset. Among other examples, large open-source clinical datasets~\cite{johnson2016mimic} have enabled researchers to release clinical contextual word embedding models. ClinicalBERT is a publicly available BERT model initialized from BioBERT~\cite{lee2020biobert} and further trained on intensive care unit notes~\cite{alsentzer2019publicly}.

\section{Dataset}

We predict IPV using a dataset of IPV victims and age-matched control patients. We predict injury using a dataset of only IPV victims, with labels from emergency radiologists.

\subsection{IPV patient selection}
The study cohort consisted of victims who were referred to a large academic hospital’s violence prevention support program between January 2013 and June 2018. For the early detection of IPV through IPV prediction, we randomly selected 265 women reporting physical abuse. We excluded all victims without any radiological studies from both groups because our algorithm seeks to predict IPV from radiology reports. The final IPV dataset consists of 262 patients.

For injury prediction, we examine a wider set of patients from two groups of victims referred to a large academic hospital’s violence prevention support program between January 2013 and June 2018. For the first group, we randomly selected 940 victims out of 2948 reporting any type of IPV-physical, psychosocial, or sexual. The second group comprised of all 308 IPV victims (including 265 women) reporting physical abuse. We excluded all victims without any radiological studies from both groups. The final IPV dataset consists of 530 patients.

\subsection{Control group selection}

We age-matched against 265 women with physical abuse and filtered for patients with at least one radiology study that was not canceled. We selected the first 795 of the resulting 1006 patients to build our control cohort. Note that the control cohort was matched against the 265 female IPV victims and does not contain any men. 

\subsection{Injury labels}
The full set of radiological studies and reports of the injury prediction patient cohort were analyzed for the presence of injury for each study. Any radiological findings unrelated to potential physical injury such as pancreatitis, malignancy, subarachnoid hemorrhage due to aneurysm rupture, etc. were not recorded as ``injury". All images were reviewed by four emergency radiology fellowship-trained radiologists who were aware of history of IPV but were blinded to the date of identification of IPV and clinical notes. The readers had full access to the radiology reports. The radiologists also recorded any injuries such as soft tissue swelling, rib fracture, etc. which might be overlooked or not mentioned in the original radiology reports. Each report was reviewed separately and labeled with an injury or not. Of the 15,639 radiology reports reviewed, 2.57\% of them were found to have an injury.

\subsection{Data cleaning}

For each radiology report, we remove extraneous information to improve clarity for the predictive models. We remove all header and footer information, punctuation, and line breaks. We change the text to lowercase and create tokens from each word through bag of words or clinicalBERT.~\cite{alsentzer2019publicly} Radiology reports that lack meaningful information after this cleaning are removed from the dataset. Patients who do not have any radiology reports after this step are removed from the dataset completely. 

\subsection{Demographic data}

We extract demographic data from IPV victims and controls including age, gender, race, and marital status. To structure free-form responses for some fields, we consolidate each field into several categories. For age, we discretize the field into $<30$, 30-50, 51-65, and 66+. The average age of patients in dataset is $43.8 \pm 18.5$, with IPV victims average age at $40.9 \pm 13.3$ and control population average age at $46.3 \pm 4.7$. For race, we consider white, Black, Hispanic, and ``other" categories with patients allowed to belong to more than one group. For marital status, we categorize single, married, and other. Note that because our control population was sex and age-matched against a cohort of female IPV victims, our control population contains no men. We do not use demographic information for predictions and use only radiology reports. For summary statistics about the dataset, see Table~\ref{tab:summary}.

\begin{table}[ht]
\tbl{Summary statistics for dataset, with percentages of radiology reports.}
{\begin{tabular}{@{}llrrrrrr@{}}
\toprule
& & \multicolumn{3}{c}{IPV Prediction} & \multicolumn{3}{c}{Injury Prediction} \\
& & Total & IPV & Control  & Total & Injury & No Injury \\ \colrule
\phantom{..}\# Patients &  & 1,056 & 262 & 794 & 530 & 135 & 395\\ 
\multicolumn{2}{l}{\# Radiology Reports} & 24,131 & 5,127 & 19,004 & 10,009 & 172 & 9,837\\ \colrule
Age & $<30$ & 6.8\% & 14.4\% & 4.8\% & 9.8\% & 10.5\% & 7.6\%\\ 
& 30-50 & 32.0\% & 47.4\% & 27.8\% & 53.6\% & 58.7\% & 58.6\%\\
& 51-65 & 37.2\% & 32.1\% & 38.5\% & 29.3\% & 21.5\% & 25.7\%\\
& 66+ & 23.9\% & 6.1\% & 28.7\% & 7.3\% & 9.3\% & 8.1\%\\ \colrule
Gender & Female &  100.0\% & 100.0\% & 100.0\% & 93.4\% & 90.7\% & 94.8\% \\
& Male & 0.0\% & 0.0\% & 0.0\% & 6.6\% & 9.3\% & 5.2\%\\ \colrule
Race & Black & 50.8\% & 34.6\% & 55.2\% & 28.3\% & 29.1\% & 23.6\% \\
& Hispanic & 12.0\% & 24.7\% & 8.6\% &22.2\% & 19.2\% & 21.9\%\\
& White & 10.0\% & 29.4\% & 4.8\% & 38.7\% & 40.7\% & 43.6\% \\
& Other & 27.6\% & 11.6\% & 32.0\% & 11.6\% & 11.6\% & 12.0\%\\ \colrule
Marital Status & Single & 45.0\% & 56.6\% & 41.8\% &50.8\% & 57.0\% & 47.9\%\\
& Married  & 36.1\% & 19.4\% & 40.7\% & 29.7\% & 27.3\% & 36.2\% \\
& Other &  18.9\% & 24.0\% & 17.5\% & 19.5\% & 15.7\% & 15.9\%\\ \botrule
\end{tabular}}\label{tab:summary}
\end{table}


\section{Methodology}
\subsection{Experiment setup}
We train our models on 60\% of the patients, validate and select hyperparameters based on 20\% of the patients, and report test performance on 20\% of the patients. To avoid data leakage, we split our data based on patient rather than radiology study. Once a patient is assigned to train, validation, or test dataset, we assign all radiology reports and labels for that patient to the corresponding dataset. We perform analysis on five trials with shuffled splits of the data. All models are compared against the same five dataset splits. 

\subsection{Models}
We compare two tasks and five models. We predict IPV and injury based on collected labels. We consider data from extracted demographic data, radiology reports, and a combination of the two. We use logistic regression, random forest, gradient boosted trees, neural network with bag of words representation, and neural network with clinicalBERT~\cite{alsentzer2019publicly} representation.

For logistic regression, we search over hyperparameters including regularization constant $C = [0.001, 0.01, 0.1, 1., 2., 5.]$ and regularization type of L1 or L2. For random forest, we search over maximum depth of trees of 10, 50, 100, 500, or no maximum depth. For gradient boosted trees, we search over hyperparameters learning rate of 0.01, 0.1, 0.5, or 1 and maximum depth of 2, 3, and 4. We use the sklearn-learn Python package~\cite{scikit-learn} with otherwise default settings.

We train two neural network models using the AllenNLP library~\cite{Gardner2017AllenNLP}. Both models contain an embedding layer followed by two feed forward layers with rectified linear unit function and linear activations. The first model represents each note as a vector of word frequencies (``Bag of Words") projected down to a lower dimensional vector while the second model leverages clinicalBERT's contextual word embeddings to represent each note. 

To facilitate more rapid training on CPUs, we freeze the clinicalBERT embeddings and only train the feed forward layers. The first model was trained for 40 epochs with an early stopping patient of 5 epochs, and the second model was trained for 10 epochs due to computational constraints. Gradient norms were rescaled to a max of 5.0, and training examples were batched by note length to minimize excess padding. Hyperparameters were selected according to validation set performance, resulting in a learning rate of 0.001, weight decay of 0.0001 and batch size of 32 for both models.

\subsection{Evaluation}

\subsubsection{Prediction and predictive features} 

We report the predictive performance as the area under the receiver operator curve (AUC) on the same train, validation, and test datasets for all models compared. We compute AUC means and standard deviations for the test datasets of the five shuffled splits of the data. 

We present predictive features by finding words with high feature importance. Because many compared models are non-linear, it is difficult to use interpretability methods to find predictive words. As logistic regression performance is comparable to that of other other non-linear methods (see Table~\ref{tab:prediction}), we present linear coefficients of the logistic regression across five test sets of the shuffled splits of the data.

\subsubsection{Error analysis} As clinical models face high stakes decisions, it is important that machine learning reduce health disparities~\cite{chen2020treating} rather than amplify existing biases.~\cite{rajkomar2018ensuring} We audit our best prediction model for IPV and injury by comparing accuracy, sensitivity, and specificity for different subgroups including age, race, gender, and marital status.~\cite{chen2019can, chen2020robustly, hardt_equality_2016} We compute means and standard deviations of performance metrics for each subgroup with model sensitivity set to 0.95 because the clinical healthcare system can accommodate many false positives---e.g. offering a conversation with a social worker---whereas false negatives can be more dire---e.g. not providing an IPV victim with additional resources for help. Predicted probabilities are computed for test datasets and compared to the true labels for the five shuffled splits of the data.

\subsubsection{Report-program date gap}
\label{sec:early_detection_setup}
One practical measure of IPV prediction is how much earlier does our model predict IPV compared to the date of patient entry into a violence prevention program. Although some patients may be reluctant to seek clinical assistance,~\cite{hien2009interpersonal} earlier detection and appropriate triage of IPV victims can help empower clinicians to intervene and provide better care for victims.~\cite{khurana2020making} For each radiology report, we compare the radiology report date with the entry date into the program. We call this difference in dates the \textit{report-program date gap}, or simply the \textit{date gap}. Negative date gaps denote reports that occur before program entry. A radiology report with a large magnitude date gap is one that occurs long before program entry whereas a low magnitude date gap occurs shortly before program entry. A model that can make predictions with a large magnitude date gap per IPV victim would allow us to allocate resources and support to high risk individuals more efficiently. For each IPV victim, we compute the largest date gap for which the model predicts IPV above the chosen threshold.

We select the prediction threshold to satisfy specificity constraints. A trivial way to maximize the early IPV detection would be to predict IPV for every patient in the dataset. This simplification would yield redundant results and a high sensitivity (true positive rate). Accordingly, we fix our specificity level (true negative rate) to be at least 95\% and compute the corresponding model threshold. We report the median earliest date gap for all IPV victims for whom the model predicts correctly.

\section{Results}

\subsection{IPV and injury prediction and predictive features}

We are able to predict IPV (best mean AUC of 0.852, random forest classifier) and injury (best mean AUC of 0.887, random forest classifier). For more results, see Table~\ref{tab:prediction}. 
We find that words that are most predictive for IPV and injury match clinical literature in IPV injury patterns from radiology reports. In Table~\ref{tab:predictive_features}, we show words with highest feature importance from logistic regression for both tasks. Findings include soft-tissue abnormalities such as swelling and hematomas and musculoskeletal injuries such as fractures. These findings reflect prior research on IPV injury patterns.~\cite{george2019radiologic}

\begin{table}[t]
\tbl{Model AUC means and standard deviations over five data splits for IPV and injury prediction using radiology reports. Bold rows indicate best performance for task.}
{\begin{tabular}{@{}lrr@{}}
\toprule
Model & IPV & Injury \\ \colrule

Logistic Regression &  0.841 $\pm$ 0.033 & 0.866 $\pm$ 0.016\\
Random Forest &  \textbf{0.852 $\pm$ 0.022} & \textbf{0.887 $\pm$ 0.019} \\
Gradient Boosted Trees &  0.842 $\pm$ 0.027 & 0.858 $\pm$ 0.030 \\
Neural Network (Bag of Words) & 0.849 $\pm$ 0.026 & 0.879 $\pm$ 0.010  \\
Neural Network (clinicalBERT~\cite{alsentzer2019publicly}) &  0.843 $\pm$ 0.022 &  0.852 $\pm$ 0.021 \\ 

\botrule

\end{tabular}}\label{tab:prediction}
\end{table}

\begin{figure}[t]
\centerline{
\includegraphics[width=0.38\textwidth]{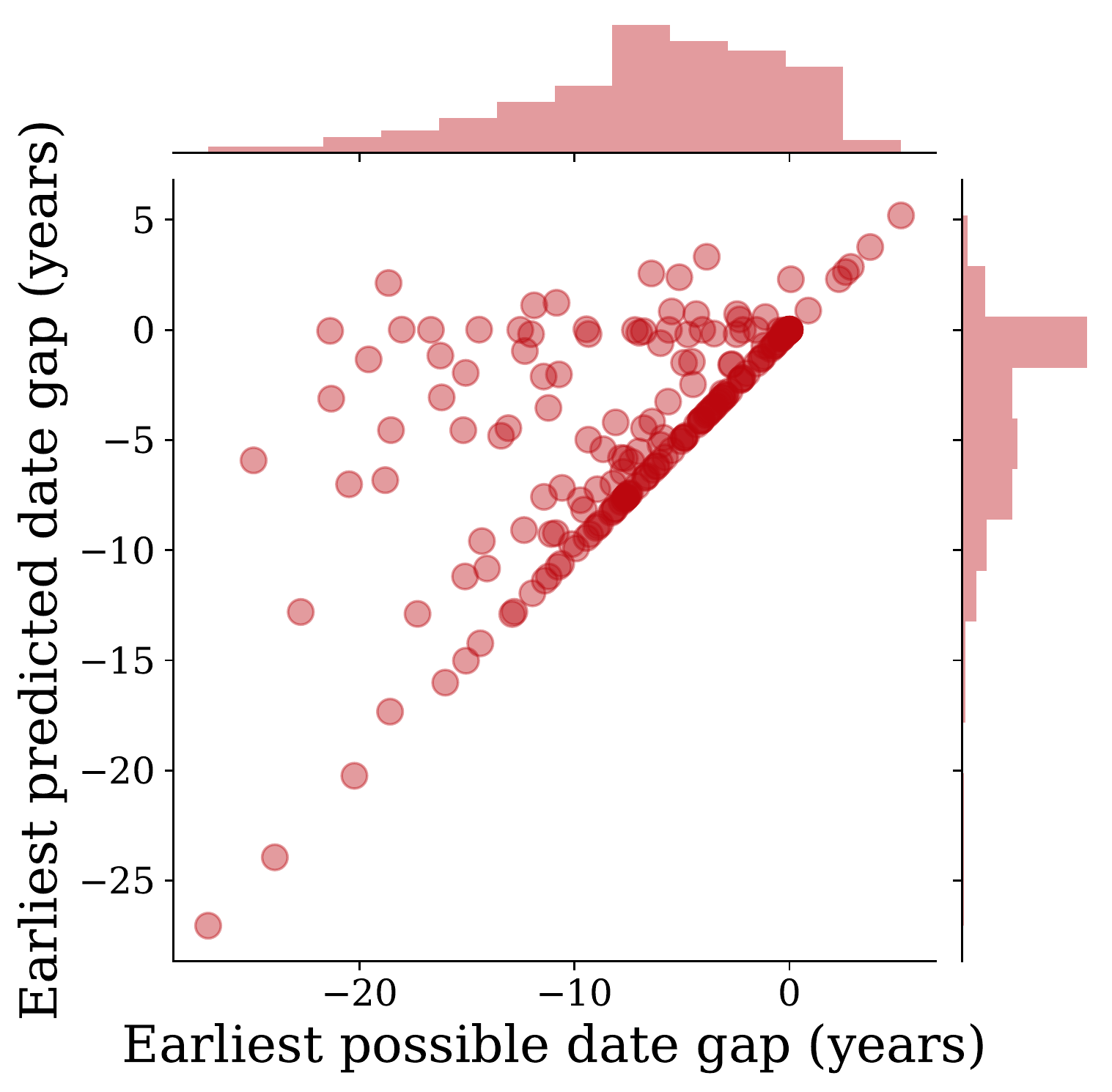}
\includegraphics[width=0.38\textwidth]{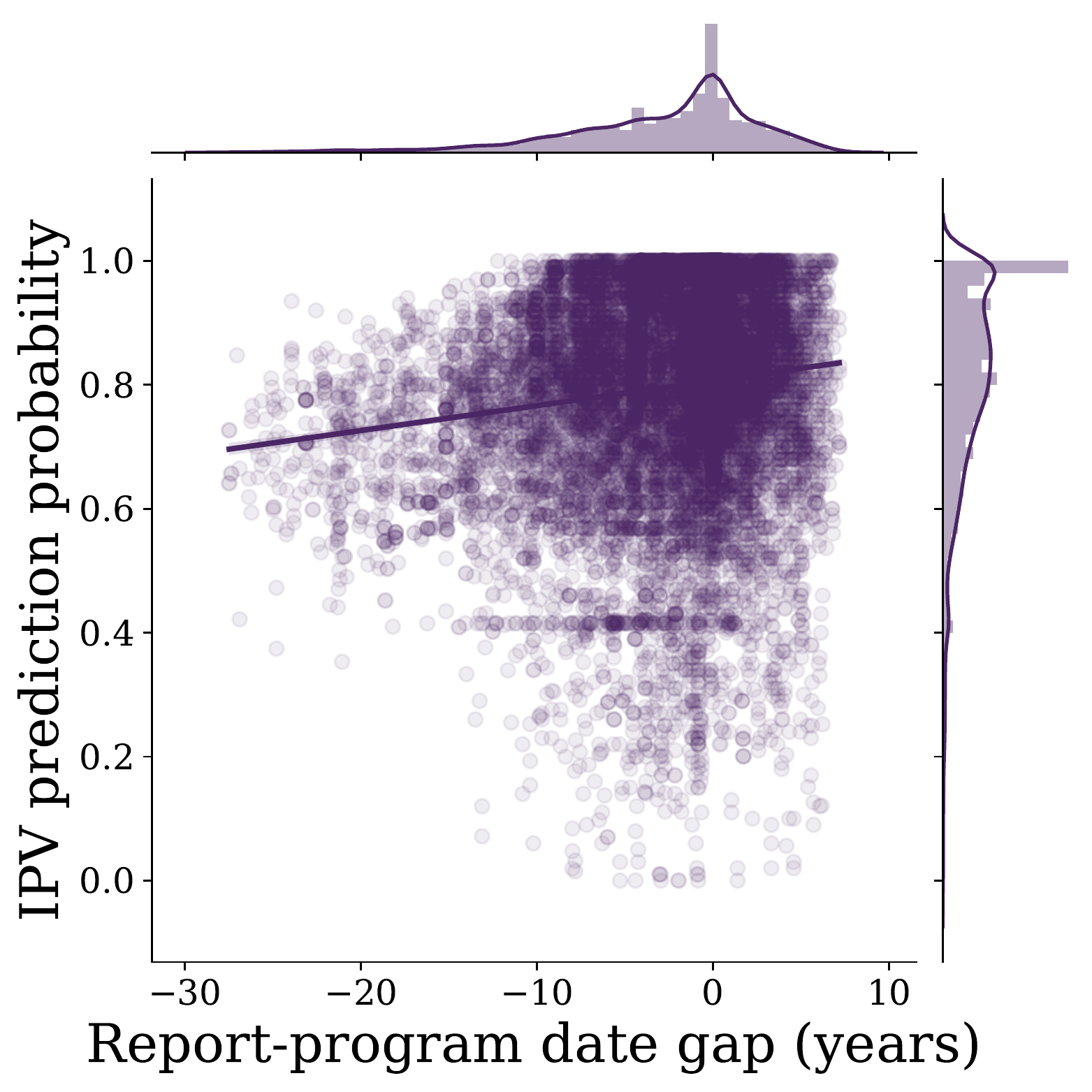}
}
\caption{Scatterplots and marginal histograms for random forest classifier for IPV prediction. \textbf{Left:} Earliest possible report-program date gap per patient ($x$-axis) compared to earliest predicted date gap ($y$-axis) with sensitivity of 64\% and specificity of 95\%. \textbf{Right:} Report-program date gap ($x$-axis) and IPV prediction probability ($y$-axis) for all radiology reports of IPV victims. }
\label{fig:detection}
\end{figure}

\begin{table}[th!]
\tbl{Predictive words for IPV and injury averaged across five trials based on linear coefficients of logistic regression. Underline indicates words consistent with clinical literature.~\cite{george2019radiologic}}
{\begin{tabular}{@{}lp{4.5in}@{}}
\toprule
Task & Predictive words \\ \colrule
IPV & ordering, final, \underline{trauma}, \underline{hematoma}, technique, \underline{swelling}, cell, \underline{fracture}, type, \underline{fractures}, lymphoma, electronically, male, pancreatitis, reason, gms, implants, unresponsive, assault, none, cancer, pregnancy, mca\\

Injury & \underline{hematoma}, \underline{fracture}, \underline{fractures}, swelling, \underline{trauma}, subchorionic, foreign, ankle, third, hand, nondisplaced, fall, stab, phalanx, finger, deformity, skullbase, fifth, wound, \underline{laceration}, sob, digit, measuring \\ 
 \botrule
\end{tabular}}\label{tab:predictive_features}
\end{table}

\subsection{Error analysis}

We find differences in performance in subgroups of age, gender, race in our error analysis (see Table~\ref{tab:error}). We focus on sensitivity because in cases of IPV and injury, it is much more important to detect all true positives. In particular, older patients (51-65, 66+) have lower sensitivity for both IPV and injury prediction. Other groups have low sensitivity for either IPV or injury prediction, but not both. For example, Black patients have lower sensitivity for injury prediction. White patients have low sensitivity for IPV prediction. It appears that patients who are not single or married (e.g. widowed, separated) have lower sensitivity for injury whereas married patients have lower sensitivity for IPV prediction.

\subsection{Report-program date gap}

We can detect IPV from radiology reports much earlier than a patient's entry into a violence prevention program. We compute the report-program date gap with specificity threshold of the random forest classifier set to 95\% and find a median date gap of 3.08 years, compared to the median earliest possible date gap of 5.83 years. For visual representations of the predicted date gap compared to the earliest possible date gap and the prediction scores, see Figure~\ref{fig:detection}. 



\begin{table}[t]
\tbl{Error analysis for IPV and injury predictions from random forest classifier. Means and standard deviations of accuracy, sensitivity (TPR), and specificity (TNR) computed over 5 data splits with overall model sensitivity set to 0.95. Bold indicates subgroups with particularly low metrics.}
{\begin{tabular}{@{}llrrrrrr@{}}
\toprule
& & \multicolumn{3}{c}{IPV Prediction} & \multicolumn{3}{c}{Injury Prediction} \\
& & Accuracy & TPR & TNR & Accuracy & TPR & TNR \\ \colrule
Age & $<30$ & 83.6 $\pm$ 4\% & 97.7 $\pm$ 1\% & 53.6 $\pm$ 11\% & 62.5 $\pm$ 11\% & 93.9 $\pm$ 9\% & 61.2 $\pm$ 11\% \\
& 30-50 &  87.4 $\pm$ 1\% & 96.3 $\pm$ 1\% & 49.6 $\pm$ 5\% & 54.1 $\pm$ 12\% & 94.8 $\pm$ 1\% & 52.9 $\pm$ 13\%\\
& 51-65 & \textbf{71.8 $\pm$ 5\%} & 92.5 $\pm$ 2\% & 49.1 $\pm$ 2\% & 41.4 $\pm$ 18\% & \textbf{89.5 $\pm$ 3\%} & 40.4 $\pm$ 19\% \\
& 66+ & \textbf{60.9 $\pm$ 5\%} & \textbf{84.4 $\pm$ 2\%} & 45.2 $\pm$ 9\% &  \textbf{33.5 $\pm$ 16\%} & 98.0 $\pm$ 4\% & 31.1 $\pm$ 17\% \\ \colrule
Gender & Female &77.2 $\pm$ 1\% & 94.6 $\pm$ 1\% & 48.4 $\pm$ 4\% &  50.0 $\pm$ 15\% & 93.4 $\pm$ 1\% & 48.9 $\pm$ 15\%\\
& Male & --- & --- & --- & \textbf{31.7 $\pm$ 21\%} & 96.2 $\pm$ 4\% & 28.8 $\pm$ 21\%\\ \colrule
Race & Black & \textbf{72.3 $\pm$ 2\%} & 95.6 $\pm$ 0\% & 41.4 $\pm$ 7\% & 47.8 $\pm$ 14\% & \textbf{88.3 $\pm$ 3\%} & 46.5 $\pm$ 14\% \\
& Hispanic & 91.1 $\pm$ 2\% & 97.9 $\pm$ 0\% & 51.9 $\pm$ 11\% & 58.0 $\pm$ 13\% & 96.9 $\pm$ 3\% & 57.4 $\pm$ 13\% \\
& White & 84.6 $\pm$ 1\% & \textbf{90.5 $\pm$ 2\%} & 43.0 $\pm$ 5\% & 41.6 $\pm$ 18\% & 95.1 $\pm$ 3\% & 39.8 $\pm$ 18\%\\
& Other & \textbf{68.7 $\pm$ 3\%} & 98.0 $\pm$ 0\% & 55.1 $\pm$ 5\%  & 58.3 $\pm$ 13\% & 95.0 $\pm$ 6\% & 57.4 $\pm$ 13\% \\ \colrule
Marital Status & Single & 81.5 $\pm$ 2\% & 95.4 $\pm$ 0\% & 45.5 $\pm$ 7\% & 49.6 $\pm$ 13\% & 95.3 $\pm$ 1\% & 48.1 $\pm$ 14\% \\
& Married  & \textbf{70.6 $\pm$ 1\%} & \textbf{92.2 $\pm$ 2\%} & 49.8 $\pm$ 3\% & 49.2 $\pm$ 18\% & 92.6 $\pm$ 7\% & 48.5 $\pm$ 19\%\\
& Other & 83.4 $\pm$ 2\% & 95.4 $\pm$ 2\% & 49.8 $\pm$ 9\% & 46.5 $\pm$ 16\% & \textbf{88.7 $\pm$ 3\%} & 45.4 $\pm$ 17\%\\ \botrule

\end{tabular}}\label{tab:error}
\end{table}

\section{Discussion and conclusion}

We present a range of findings on the use of prediction algorithms to address IPV in the clinical setting through the analysis of radiology reports. Our results demonstrate several main takeaways. First, with a dataset of 34,642 reports and 1,479 patients, we are able accurately predict IPV and injury with AUCs of 0.852 and 0.887, respectively. Second, while our algorithm demonstrates some bias in the form of differences in accuracy, sensitivity, and specificity with respect to age, gender, race, and marital status, we are able to predict a median report-program date gap of over 3.08 years with sensitivity of 64\% and specificity of 95\%. 

Our work leads naturally to many directions for future research. One limitation of our current work is that we consider one radiology report at a time for IPV and injury prediction and exclude clinical history. Because IPV victims seek greater medical care from clinical settings like the emergency department,~\cite{wisner1999intimate, dearwater1998prevalence} patient data including previous visits, clinical notes, and diagnoses could yield more accurate predictions and therefore earlier detection.~\cite{khurana2020making} Additionally, predictive algorithms can help identify the best intervention for an IPV victim. Currently screening programs for IPV vary in execution and effect,~\cite{o2015screening} and once screened, IPV victims face many obstacles before leaving an abusive relationship.~\cite{kim2008leave} Deeper understanding of targeted interventions could provide a crucial contribution to patient advocacy.

Deployment of a predictive model for IPV and injury detection faces several practical challenges. As with many machine learning algorithms in clinical settings, question of generalization across hospitals~\cite{ghassemi2019practical} and across subgroups~\cite{chen2019can} raise concerns about robustness and fairness. Moreover, better understanding of physician reliance on, distrust of, and confusion towards predictive models in clinical settings is an active area of research.~\cite{tschandl2020human} We have shown in our analysis that automated detection through machine learning can predict IPV and injury from radiology reports. We look forward to future work towards the deployment of an IPV early detection model in a clinical setting.



\bibliographystyle{ws-procs11x85}
\bibliography{references}

\begin{thebibliography}{10}

\bibitem{fulu2013prevalence}
E.~Fulu, R.~Jewkes, T.~Roselli, C.~Garcia-Moreno {\em et~al.}, Prevalence of
  and factors associated with male perpetration of intimate partner violence:
  findings from the un multi-country cross-sectional study on men and violence
  in asia and the pacific, {\em The lancet global health} {\bf 1}, e187
  (2013).

\bibitem{black2011national}
M.~Black, K.~Basile, M.~Breiding, S.~Smith, M.~Walters, M.~Merrick, J.~Chen and
  M.~Stevens, National intimate partner and sexual violence survey: 2010
  summary report  (2011).

\bibitem{campbell2002health}
J.~C. Campbell, Health consequences of intimate partner violence, {\em The
  lancet} {\bf 359}, 1331  (2002).

\bibitem{tollestrup1999health}
K.~Tollestrup, D.~Sklar, F.~J. Frost, L.~Olson, J.~Weybright, J.~Sandvig and
  M.~Larson, Health indicators and intimate partner violence among women who
  are members of a managed care organization, {\em Preventive medicine} {\bf
  29}, 431  (1999).

\bibitem{ellsberg2008intimate}
M.~Ellsberg, H.~A. Jansen, L.~Heise, C.~H. Watts, C.~Garcia-Moreno {\em
  et~al.}, Intimate partner violence and women's physical and mental health in
  the who multi-country study on women's health and domestic violence: an
  observational study, {\em The lancet} {\bf 371}, 1165  (2008).

\bibitem{united2018global}
U.~N.~O. on~Drugs and Crime, {\em Global Study on Homicide: Gender-related
  Killing of Women and Girls} (UNODC, United Nations Office on Drugs and Crime,
  2018).

\bibitem{wisner1999intimate}
C.~Wisner, T.~Gilmer, L.~Saltzman and T.~Zink, Intimate partner violence
  against women, {\em Journal of family practice} {\bf 48}, 439  (1999).

\bibitem{dearwater1998prevalence}
S.~R. Dearwater, J.~H. Coben, J.~C. Campbell, G.~Nah, N.~Glass, E.~McLoughlin
  and B.~Bekemeier, Prevalence of intimate partner abuse in women treated at
  community hospital emergency departments, {\em Jama} {\bf 280}, 433  (1998).

\bibitem{russo2019imaging}
A.~Russo, A.~Reginelli, M.~Pignatiello, F.~Cioce, G.~Mazzei, O.~Fabozzi,
  V.~Parlato, S.~Cappabianca and S.~Giovine, Imaging of violence against the
  elderly and the women,  {\bf 40}, 18  (2019).

\bibitem{george2019radiologic}
E.~George, C.~H. Phillips, N.~Shah, A.~Lewis-O’Connor, B.~Rosner, H.~M.
  Stoklosa and B.~Khurana, Radiologic findings in intimate partner violence,
  {\em Radiology} {\bf 291}, 62  (2019).

\bibitem{lipsky2006role}
S.~Lipsky, R.~Caetano, C.~A. Field and G.~L. Larkin, The role of intimate
  partner violence, race, and ethnicity in help-seeking behaviors, {\em
  Ethnicity and Health} {\bf 11}, 81  (2006).

\bibitem{tjaden2000prevalence}
P.~Tjaden and N.~Thoennes, Prevalence and consequences of male-to-female and
  female-to-male intimate partner violence as measured by the national violence
  against women survey, {\em Violence against women} {\bf 6}, 142  (2000).

\bibitem{rennison2001intimate}
C.~M. Rennison, {\em Intimate partner violence and age of victim, 1993-99} (US
  Department of Justice, Office of Justice Programs, Bureau of Justice~…,
  2001).

\bibitem{salomon2002relationship}
A.~Salomon, S.~S. Bassuk and N.~Huntington, The relationship between intimate
  partner violence and the use of addictive substances in poor and homeless
  single mothers, {\em Violence Against Women} {\bf 8}, 785  (2002).

\bibitem{van2020covid}
N.~Van~Gelder, A.~Peterman, A.~Potts, M.~O'Donnell, K.~Thompson, N.~Shah and
  S.~Oertelt-Prigione, Covid-19: Reducing the risk of infection might increase
  the risk of intimate partner violence, {\em EClinicalMedicine} {\bf 21}
  (2020).

\bibitem{gosangi2020exacerbation}
B.~Gosangi, H.~Park, R.~Thomas, R.~Gujrathi, C.~P. Bay, A.~S. Raja, S.~E.
  Seltzer, M.~C. Balcom, M.~L. McDonald, D.~P. Orgill {\em et~al.},
  Exacerbation of physical intimate partner violence during covid-19 lockdown,
  {\em Radiology} , p. 202866  (2020).

\bibitem{stockl2013global}
H.~St{\"o}ckl, K.~Devries, A.~Rotstein, N.~Abrahams, J.~Campbell, C.~Watts and
  C.~G. Moreno, The global prevalence of intimate partner homicide: a
  systematic review, {\em The Lancet} {\bf 382}, 859  (2013).

\bibitem{hien2009interpersonal}
D.~Hien and L.~Ruglass, Interpersonal partner violence and women in the united
  states: An overview of prevalence rates, psychiatric correlates and
  consequences and barriers to help seeking, {\em International journal of law
  and psychiatry} {\bf 32}, p.~48  (2009).

\bibitem{khurana2020making}
B.~Khurana, S.~E. Seltzer, I.~S. Kohane and G.~W. Boland, Making the
  ‘invisible’visible: transforming the detection of intimate partner
  violence, {\em BMJ quality \& safety} {\bf 29}, 241  (2020).

\bibitem{kraanen2014prediction}
F.~L. Kraanen, E.~Vedel, A.~Scholing and P.~M. Emmelkamp, Prediction of
  intimate partner violence by type of substance use disorder, {\em Journal of
  substance abuse treatment} {\bf 46}, 532  (2014).

\bibitem{ghassemi2020review}
M.~Ghassemi, T.~Naumann, P.~Schulam, A.~L. Beam, I.~Y. Chen and R.~Ranganath, A
  review of challenges and opportunities in machine learning for health, {\em
  AMIA Summits on Translational Science Proceedings} {\bf 2020}, p. 191
  (2020).

\bibitem{ghassemi2019practical}
M.~Ghassemi, T.~Naumann, P.~Schulam, A.~L. Beam, I.~Y. Chen and R.~Ranganath,
  Practical guidance on artificial intelligence for health-care data, {\em The
  Lancet Digital Health} {\bf 1}, e157  (2019).

\bibitem{razavian2015population}
N.~Razavian, S.~Blecker, A.~M. Schmidt, A.~Smith-McLallen, S.~Nigam and
  D.~Sontag, Population-level prediction of type 2 diabetes from claims data
  and analysis of risk factors, {\em Big Data} {\bf 3}, 277  (2015).

\bibitem{kehl2019assessment}
K.~L. Kehl, H.~Elmarakeby, M.~Nishino, E.~M. Van~Allen, E.~M. Lepisto, M.~J.
  Hassett, B.~E. Johnson and D.~Schrag, Assessment of deep natural language
  processing in ascertaining oncologic outcomes from radiology reports, {\em
  JAMA oncology} {\bf 5}, 1421  (2019).

\bibitem{saria2010integration}
S.~Saria, A.~K. Rajani, J.~Gould, D.~Koller and A.~A. Penn, Integration of
  early physiological responses predicts later illness severity in preterm
  infants, {\em Science translational medicine} {\bf 2}, 48ra65  (2010).

\bibitem{chouldechova2018case}
A.~Chouldechova, D.~Benavides-Prado, O.~Fialko and R.~Vaithianathan, A case
  study of algorithm-assisted decision making in child maltreatment hotline
  screening decisions, 134  (2018).

\bibitem{brown2019toward}
A.~Brown, A.~Chouldechova, E.~Putnam-Hornstein, A.~Tobin and R.~Vaithianathan,
  Toward algorithmic accountability in public services: A qualitative study of
  affected community perspectives on algorithmic decision-making in child
  welfare services, 1  (2019).

\bibitem{wu2020deep}
S.~Wu, K.~Roberts, S.~Datta, J.~Du, Z.~Ji, Y.~Si, S.~Soni, Q.~Wang, Q.~Wei,
  Y.~Xiang {\em et~al.}, Deep learning in clinical natural language processing:
  a methodical review, {\em Journal of the American Medical Informatics
  Association} {\bf 27}, 457  (2020).

\bibitem{gao2018hierarchical}
S.~Gao, M.~T. Young, J.~X. Qiu, H.-J. Yoon, J.~B. Christian, P.~A. Fearn, G.~D.
  Tourassi and A.~Ramanthan, Hierarchical attention networks for information
  extraction from cancer pathology reports, {\em Journal of the American
  Medical Informatics Association} {\bf 25}, 321  (2018).

\bibitem{chen2019clinical}
L.~Chen, Y.~Gu, X.~Ji, C.~Lou, Z.~Sun, H.~Li, Y.~Gao and Y.~Huang, Clinical
  trial cohort selection based on multi-level rule-based natural language
  processing system, {\em Journal of the American Medical Informatics
  Association} {\bf 26}, 1218  (2019).

\bibitem{afzal2018natural}
N.~Afzal, V.~P. Mallipeddi, S.~Sohn, H.~Liu, R.~Chaudhry, C.~G. Scott, I.~J.
  Kullo and A.~M. Arruda-Olson, Natural language processing of clinical notes
  for identification of critical limb ischemia, {\em International journal of
  medical informatics} {\bf 111}, 83  (2018).

\bibitem{poulin2014predicting}
C.~Poulin, B.~Shiner, P.~Thompson, L.~Vepstas, Y.~Young-Xu, B.~Goertzel,
  B.~Watts, L.~Flashman and T.~McAllister, Predicting the risk of suicide by
  analyzing the text of clinical notes, {\em PloS one} {\bf 9}, p. e85733
  (2014).

\bibitem{huang2019clinicalbert}
K.~Huang, J.~Altosaar and R.~Ranganath, Clinicalbert: Modeling clinical notes
  and predicting hospital readmission, {\em arXiv preprint arXiv:1904.05342}
  (2019).

\bibitem{rumshisky2016predicting}
A.~Rumshisky, M.~Ghassemi, T.~Naumann, P.~Szolovits, V.~Castro, T.~McCoy and
  R.~Perlis, Predicting early psychiatric readmission with natural language
  processing of narrative discharge summaries, {\em Translational psychiatry}
  {\bf 6}, e921  (2016).

\bibitem{pham2014natural}
A.-D. Pham, A.~N{\'e}v{\'e}ol, T.~Lavergne, D.~Yasunaga, O.~Cl{\'e}ment,
  G.~Meyer, R.~Morello and A.~Burgun, Natural language processing of radiology
  reports for the detection of thromboembolic diseases and clinically relevant
  incidental findings, {\em BMC bioinformatics} {\bf 15}, 1  (2014).

\bibitem{boag2018s}
W.~Boag, D.~Doss, T.~Naumann and P.~Szolovits, What’s in a note? unpacking
  predictive value in clinical note representations, {\em AMIA Summits on
  Translational Science Proceedings} {\bf 2018}, p.~26  (2018).

\bibitem{devlin2018bert}
J.~Devlin, M.-W. Chang, K.~Lee and K.~Toutanova, Bert: Pre-training of deep
  bidirectional transformers for language understanding, {\em arXiv preprint
  arXiv:1810.04805}   (2018).

\bibitem{liu2019roberta}
Y.~Liu, M.~Ott, N.~Goyal, J.~Du, M.~Joshi, D.~Chen, O.~Levy, M.~Lewis,
  L.~Zettlemoyer and V.~Stoyanov, Roberta: A robustly optimized bert
  pretraining approach, {\em arXiv preprint arXiv:1907.11692}   (2019).

\bibitem{lan2019albert}
Z.~Lan, M.~Chen, S.~Goodman, K.~Gimpel, P.~Sharma and R.~Soricut, Albert: A
  lite bert for self-supervised learning of language representations, {\em
  arXiv preprint arXiv:1909.11942}   (2019).

\bibitem{johnson2016mimic}
A.~E. Johnson, T.~J. Pollard, L.~Shen, H.~L. Li-Wei, M.~Feng, M.~Ghassemi,
  B.~Moody, P.~Szolovits, L.~A. Celi and R.~G. Mark, Mimic-iii, a freely
  accessible critical care database, {\em Scientific data} {\bf 3}, 1  (2016).

\bibitem{lee2020biobert}
J.~Lee, W.~Yoon, S.~Kim, D.~Kim, S.~Kim, C.~H. So and J.~Kang, Biobert: a
  pre-trained biomedical language representation model for biomedical text
  mining, {\em Bioinformatics} {\bf 36}, 1234  (2020).

\bibitem{alsentzer2019publicly}
E.~Alsentzer, J.~R. Murphy, W.~Boag, W.-H. Weng, D.~Jin, T.~Naumann and
  M.~McDermott, Publicly available clinical bert embeddings, {\em arXiv
  preprint arXiv:1904.03323} {\bf 2019}  (2019).

\bibitem{scikit-learn}
F.~Pedregosa, G.~Varoquaux, A.~Gramfort, V.~Michel, B.~Thirion, O.~Grisel,
  M.~Blondel, P.~Prettenhofer, R.~Weiss, V.~Dubourg, J.~Vanderplas, A.~Passos,
  D.~Cournapeau, M.~Brucher, M.~Perrot and E.~Duchesnay, Scikit-learn: Machine
  learning in {P}ython, {\em Journal of Machine Learning Research} {\bf 12},
  2825  (2011).

\bibitem{Gardner2017AllenNLP}
M.~Gardner, J.~Grus, M.~Neumann, O.~Tafjord, P.~Dasigi, N.~F. Liu, M.~Peters,
  M.~Schmitz and L.~S. Zettlemoyer, Allennlp: A deep semantic natural language
  processing platform (March 2017).

\bibitem{chen2020treating}
I.~Y. Chen, S.~Joshi and M.~Ghassemi, Treating health disparities with
  artificial intelligence, {\em Nature Medicine} {\bf 26}, 16  (2020).

\bibitem{rajkomar2018ensuring}
A.~Rajkomar, M.~Hardt, M.~D. Howell, G.~Corrado and M.~H. Chin, Ensuring
  fairness in machine learning to advance health equity, {\em Annals of
  internal medicine} {\bf 169}, 866  (2018).

\bibitem{chen2019can}
I.~Y. Chen, P.~Szolovits and M.~Ghassemi, Can ai help reduce disparities in
  general medical and mental health care?, {\em AMA journal of ethics} {\bf
  21}, 167  (2019).

\bibitem{chen2020robustly}
I.~Y. Chen, M.~Agrawal, S.~Horng and D.~Sontag, Robustly extracting medical
  knowledge from ehrs: A case study of learning a health knowledge graph, 19
  (January 2020).

\bibitem{hardt_equality_2016}
M.~Hardt, E.~Price and N.~Srebro, Equality of {Opportunity} in {Supervised}
  {Learning}, 3323 (June 2016), Barcelona, Spain.

\bibitem{o2015screening}
L.~O'Doherty, K.~Hegarty, J.~Ramsay, L.~L. Davidson, G.~Feder and A.~Taft,
  Screening women for intimate partner violence in healthcare settings, {\em
  Cochrane database of systematic reviews}   (2015).

\bibitem{kim2008leave}
J.~Kim and K.~A. Gray, Leave or stay? battered women's decision after intimate
  partner violence, {\em Journal of Interpersonal Violence} {\bf 23}, 1465
  (2008).

\bibitem{tschandl2020human}
P.~Tschandl, C.~Rinner, Z.~Apalla, G.~Argenziano, N.~Codella, A.~Halpern,
  M.~Janda, A.~Lallas, C.~Longo, J.~Malvehy {\em et~al.}, Human--computer
  collaboration for skin cancer recognition, {\em Nature Medicine} , 1  (2020).

\end{thebibliography}

\end{document}